\documentclass[authorversion,sigconf,screen]{acmart} 

\acmSubmissionID{v2rtp4198}

\usepackage{subcaption}
\usepackage[skip=3pt]{caption}
\usepackage{booktabs}
\usepackage{graphicx}
\usepackage[normalem]{ulem}
\useunder{\uline}{\ul}{}
\usepackage{amsthm}
\usepackage[inline]{enumitem}
\usepackage[acronym]{glossaries}
\makeglossaries
\usepackage[textsize=tiny]{todonotes}
\usepackage{multirow}
\usepackage[capitalize,noabbrev]{cleveref}
\usepackage{amsmath}
\usepackage{makecell}
\usepackage{array}
\usepackage{algorithm}
\usepackage{algpseudocode}
\usepackage{balance}

\theoremstyle{definition}

\AtBeginDocument{%
  \providecommand\BibTeX{{%
    \normalfont B\kern-0.5em{\scshape i\kern-0.25em b}\kern-0.8em\TeX}}}

\newacronym{pgm}{PGM}{Probabilistic Graphical Model}

\newacronym{nn}{NN}{Neural Network}
\newacronym{llm}{LLM}{Large language model}
\newacronym{serp}{SERP}{Search Engine Results Page}
\newacronym{pbm}{PBM}{Position-Based Model}
\newacronym{cm}{CM}{Cascade Model}
\newacronym{ubm}{UBM}{User Browsing Model}
\newacronym{dcm}{DCM}{Dependent Click Model}
\newacronym{dbn}{DBN}{Dynamic Bayesian Network Model}
\newacronym{ncm}{NCM}{Neural Click Model}
\newacronym{csm}{CSM}{Click Sequence Model}
\newacronym{pcsm}{PCSM}{Partially Sequential Click Model}
\newacronym{tcm}{TCM}{Temporal Click Model}
\newacronym{thcm}{THCM}{Temporal Hidden Click Model}
\newacronym{emalgorithm}{EM}{Expectation-Maximization}
\newacronym{ips}{IPS}{Inverse Propensity Scoring}
\newacronym{mle}{MLE}{Maximum Likelihood Estimation}

\makeatletter
\newcommand\headingnodot{\def\@toclevel{4}%
  \@startsection{paragraph}{4}{\z@}%
  {-.2\baselineskip \@plus -2\p@ \@minus -.2\p@}%
  {-3.5\p@}%
  {\ACM@NRadjust{\bfseries}}}
\newcommand{\heading}[1]{\headingnodot{#1.}}
\makeatother

\author{Santiago de Leon-Martinez}
\affiliation{%
  \institution{Brno University of Technology}
  \city{Brno}
  \country{Czechia}
}
\additionalaffiliation{%
  \institution{Kempelen Institute of Intelligent Technologies}
  \city{Bratislava}
  \country{Slovakia}
}
\email{santiago.deleon@kinit.sk}
\orcid{0000-0002-2109-9420}

\author{Robert Moro}
\affiliation{%
  \institution{Kempelen Institute of Intelligent Technologies}
  \city{Bratislava}
  \country{Slovakia}
}
\email{robert.moro@kinit.sk}
\orcid{0000-0002-3052-8290}

\author{Branislav Kveton}
\orcid{0000-0002-3965-1367}
\affiliation{%
  \institution{Adobe Research}
  \city{San Jose, CA}
  \country{United States}
}
\email{kveton@adobe.com}

\author{Maria Bielikova}
\orcid{0000-0003-4105-3494}
\affiliation{%
  \institution{Kempelen Institute of Intelligent Technologies}
  \city{Bratislava}
  \country{Slovakia}
}
\email{maria.bielikova@kinit.sk}

\copyrightyear{2026}
\acmYear{2026}
\setcopyright{cc}
\setcctype{by}
\acmConference[KDD '26]{Proceedings of the 32nd ACM SIGKDD Conference on Knowledge Discovery and Data Mining V.2}{August 09--13, 2026}{Jeju Island, Republic of Korea}
\acmBooktitle{Proceedings of the 32nd ACM SIGKDD Conference on Knowledge Discovery and Data Mining V.2 (KDD '26), August 09--13, 2026, Jeju Island, Republic of Korea}
\acmDOI{10.1145/3770855.3818211}
\acmISBN{979-8-4007-2259-2/2026/08}



\begin{document}

\title[From Latent to Observable Position-Based Click Models in Carousels]{From Latent to Observable Position-Based Click Models in Carousel Interfaces}

\begin{abstract}
Click models are a central component of learning and evaluation in recommender systems, yet most existing models are designed for single ranked list interfaces. In contrast, modern recommender platforms increasingly use complex interfaces, such as carousels, which consist of multiple swipeable lists that enable complex user browsing behaviors. 

In this paper, we study position-based click models in carousel interfaces and examine optimization methods, model structure, and alignment with user behavior. We propose three novel position-based models tailored to carousels, including the first position-based model without latent variables that incorporates observed examination signals derived from eye tracking data, called the Observed Examination Position-Based Model (OEPBM). We develop a general implementation of these carousel click models, supporting multiple optimization techniques and conduct experiments comparing gradient-based methods with classic approaches, namely expectation-maximization and maximum likelihood estimation.

Our results show that gradient-based optimization consistently achieves better click likelihoods. Among the evaluated models, the OEPBM  achieves the strongest performance in click prediction and produces examination patterns that most closely align to user behavior. However, we also demonstrate that strong click fit does not imply realistic modeling of user examination and browsing patterns. This reveals a fundamental limitation of click-only models in complex interfaces and the need for incorporating additional behavioral signals when designing click models for carousel-based recommender systems.

\end{abstract}

\maketitle
\newcommand\kddavailabilityurl{https://zenodo.org/records/20508190}
\ifdefempty{\kddavailabilityurl}{}{
\begingroup\small\noindent\raggedright\textbf{Resource Availability:}\\
% please change the following context to include multiple artifacts if necessary, including data, models, code, etc.
The dataset of this paper has been made publicly available at \url{https://doi.org/10.5281/zenodo.20508189}.
\endgroup
}

\section{Introduction}
Click models are probabilistic user models that infer relevance signals, biases, and user behavior from observed click interactions. A variety of click models has been proposed to capture different assumptions about user behavior and examination patterns \cite{10.1145/1526709.1526711,10.1145/1526709.1526712, 10.1145/2872427.2883033, 10.1145/3209978.3210004, 10.1145/1390334.1390392, 10.1145/1498759.1498818}. However, almost all existing clicks models are designed for single ranked list interfaces and are typically trained using classic methods like expectation-maximization (EM) or maximum likelihood estimation (MLE) \cite{chuklin_markov_rijke_click_models_2015}. At the same time, modern recommender systems increasingly rely on gradient-based optimization and are deployed in much more complex multi-list interfaces.

A prominent example of a multi-list interface is the carousel interface, which has been widely adopted by e-commerce and streaming services like Netflix and Spotify (see \cref{fig:user_study}). Carousel interfaces consist of multiple lists or \textit{carousels} that are \textit{swipeable} (i.e. can be scrolled horizontally). Similar to a grid, this enables users to browse both vertically across lists and horizontally within a list. However, they distinguish themselves from grids by each carousel being a distinct, separate list with a topic that represents the items in the carousel, such as \textit{Gems for You.} Prior work suggests that carousel interfaces increase perceived diversity and novelty \cite{10.1145/3450613.3456809,starke_serving_2021,loepp_how_2023},  improve choice satisfaction~\cite{starke_serving_2021}, and enable more efficient search behavior \cite{rahdari_towards_2024}. Despite their prevalence, user modeling for carousel interfaces remains relatively underexplored.

Recent work has begun to address this gap, including the first carousel click model \cite{rahdari_towards_2024}, the release of a carousel interaction dataset \cite{RecGaze}, the first eye tracking study of carousel browsing behavior \cite{10.1145/3742413.3789166}, and a theoretical taxonomy for carousel click models \cite{rethinking}. However, key work is lacking in computational or algorithmic methods (e.g, click models or learning to rank) that leverage the context of carousel interfaces to improve recommendations. In this paper, we aim to further develop the field of interface-aware recommender systems by designing, implementing, and testing carousel click models that not only capture clicks, but model realistic user examination and browsing behavior. We answer the following research questions: 
\begin{enumerate}
    \item[RQ~1] Do gradient-based optimization methods outperform EM and MLE when fitting click models for carousels?
    \item[RQ~2] Do carousel position-based models better capture click behavior in carousels than the existing cascade-based models?  
    \item[RQ~3] Do carousel position-based models capture realistic user examination? 
\end{enumerate}

In answering these questions, we make the following contributions:
 \begin{itemize}
    \item We propose and formally define three novel carousel position-based click models, including the first position-based click model without latent variables that incorporates observed examination signals derived from eye tracking data, termed the Observed Examination Position-Based Model (OEPBM).
    \item We develop a general implementation of these models and the two existing cascade-based models for carousels, supporting multiple optimization techniques. 
    \item We conduct empirical evaluations comparing optimization methods and click models, assessing both click prediction performance and the capability of models to correctly reflect user examination behavior, with evidence of the OEPBM outperforming other models in both clicks and modeling realistic user examination behavior.
    \item We show that, in complex interfaces such as carousels, optimizing for click likelihood alone does not guarantee realistic user behavior modeling, revealing a fundamental limitation shared by click-only models that motivates the incorporation of additional behavioral signals.
\end{itemize}

\section{Related Works}
\heading{Click Models for Single Lists}
Click models have been extensively studied as a means of inferring relevance and simulating clicks using interaction data in retrieval and recommendation settings \cite{10.1145/1526709.1526711,10.1145/1526709.1526712, 10.1145/2872427.2883033, 10.1145/3209978.3210004, 10.1145/1390334.1390392, 10.1145/1498759.1498818}. Two of the most influential and popular click models are the Cascade Model (CM) \cite{craswell_experimental_2008} and the Position Based Model (PBM) \cite{PBM}. The CM assumes a sequential browsing process, in which users examine results from top to bottom and decide at each rank whether to click or move on the next item. It is fairly simple and intuitive, but imposes a strong sequential browsing assumption. Conversely, the PBM has no sequential browsing assumption and models examination as ranked position probability independent of other positions and items. Under the PBM, a user clicks an attractive item if it was examined, while in the CM, a user clicks an attractive item only after seeing all previous items and deeming them unattractive. Thus, the PBM is much more flexible, but requires learning examination probabilities for each position, which can be difficult as it must be modeled latently without any observable signal (unlike observable clicks for attraction). The classic approach to fitting the CM, with solely observable attraction, is the maximum likelihood estimation, while for the PBM, with observable attraction and latent examination, expectation-maximization is used \cite{chuklin_markov_rijke_click_models_2015}. A recent click modeling work has shown that gradient based optimization techniques  have many advantages over these classic methods \cite{hager2025claxfastflexibleneural}. Much like their optimization methods, the PBM and CM are outdated as they cannot account for more complex multi-list interfaces like carousels.

\heading{Click Models for Carousels}
\citet{rahdari_towards_2024,rahdari_ranked_2022} proposed the first and only click model designed for carousels called the Carousel Click Model (CCM) that extends the cascade model to carousels. The model assumes that users browse carousel row topics from the top until finding one of interest and then only browse items in that carousel row with a left-right cascade behavior until finding an item of interest, while additionally adding the possibility of the user to leave unsatisfied. However, results from an eye tracked user study of carousel interfaces for movie recommendation \cite{RecGaze} and follow-up analyses \cite{10.1145/3742413.3789166} showed that users often scanned items in multiple carousels rows and scanned the initial few items of a row before moving on to the next. Moreover, the general browsing behavior (and click behavior) was an F-pattern where users are more likely to examine the top carousels and explore deeper into them. 
However, on swiping, eye tracking showed this behavior became mirrored (from left-right to right-left) as users began with the farthest right item \cite{10.1145/3742413.3789166}. Based on these eye tracking results, the CCM does not align with user behavior. Works have proposed that a PBM adapted to carousels would be able to better model behavior in carousels \cite{rahdari_towards_2024,wsdm}. The adaptation of the PBM to carousels along with a recent taxonomy of carousel click models \cite{rethinking} classifying models by observable variables, provided the motivation for this work. We explore the adaption of the PBM to carousels and design three PBM variants, one of which transforms the latent variable of examination to an observable variable using eye tracking data. Additionally, we also implement them with gradient based optimization methods to be in line with the state of the art \cite{hager2025claxfastflexibleneural}.

\section{Background: Carousel Cascade Models}
In this section, we provide a description and generalized formulation of previous click models for carousel interfaces.  We use the formulation of the PBM to describe all prior and novel models:
\begin{multline}
P(click \mid position, \: item ) = \\
P(examination \mid position) \times P(attraction \mid item)  
\end{multline}
where $P(examination \mid position)$ is the probability of examining a position and $P(attraction \mid item)$ is the probability that an item is attractive and the two probabilities are independent of each other. %and use the following notation to define a carousel. 

Assume we have a carousel interface with $I$ carousel rows and $J$ item columns per carousel. Let $i \in [I]$ represents a carousel row position and $j \in [J]$ represent the item column position in a carousel. Let $\theta_{u}$ be the attraction probability for item $u$. Therefore, our general formulation for carousel click models is:
\begin{equation}
\label{eq:Formulation General Carousel PBM}
P(click \mid i,j, u) = P(examination \mid i,j )\times \theta_{u} 
\end{equation}
Many click models differ only in the examination term. The problem lies in modifying the examination term to be able to fit click data and user behavior in carousels. \citet{rahdari_ranked_2022,rahdari_towards_2024} was the first to do this by extending cascade models to carousels interfaces.

\heading{Cascade Model for Carousels} The cascade model (CM) for single ranked lists \cite{craswell_experimental_2008} uses a cascade browsing assumption for examination: (1) the first position is always examined; and (2) any subsequent examination probabilities are the probability that the previous items were not clicked.  \citet{rahdari_ranked_2022,rahdari_towards_2024}  extended this to carousels by transforming a carousel into a single ranked list. The first row of the carousel is treated as the first 1 through $J$ items while the second carousel row are the $J+1$ through $2J$ items and so on. To simplify notation for the CM, we will use $\theta_{i',j'}$ to refer the attraction probability of the previous item at position $i',j'$.
The formulation for the CM for carousels follows:
\begin{equation}
P(click \mid i,j, u) = \prod_{i^{\prime}=1}^{i-1}\prod_{j^{\prime}=1}^{J}(1-\theta_{i^{\prime},j^{\prime}}) \prod_{j^{ \prime}=1}^{j-1}(1-\theta_{i,j^{\prime}}) \times \theta_{u} 
\end{equation}
Thus, the examination probability transforms into two product terms representing the probability of not clicking previous items. The first term represents all unattractive items of previous rows, while the second represents unattractive items of the current row. 

\heading{Terminating Cascade Model (TCM)} The TCM \cite{rahdari_ranked_2022,rahdari_towards_2024} is an extension of the CM that adds a probability to terminate after every unattractive item to account for the user stopping their search due to dissatisfaction with already seen items. 
The formulation of the TCM for carousels is:
\begin{equation}
P(click \mid i,j, u)=
(1-\kappa)^{(i-1)J+j-1}\prod_{i^{\prime}=1}^{i-1}\prod_{j^{\prime}=1}^{J}(1-\theta_{i^{\prime},j^{\prime}}) \prod_{j^{ \prime}=1}^{j-1}(1-\theta_{i,j^{\prime}}) \times \theta_{u} 
\end{equation}
where $\kappa$ is a single, fixed termination probability that the user will leave after examining an unattractive item. There is one $1-\kappa$ for every previous position and unattractive item, a total of $(i-1)J+j-1$.  The TCM was a baseline for the \textit{Carousel Click Model} (CCM). As the only two prior click models for carousels, both the CCM and TCM, will serve as baselines for the experiments below.

\heading{Carousel Click Model (CCM)}  The CCM \cite{rahdari_ranked_2022,rahdari_towards_2024} is an extension of the TCM that models a user that: (1) starts from the first carousel row topic and examines topics until finding one that is attractive, (2) if a topic is unattractive then all items in that row/topic are unattractive and there is a termination probability $\kappa_{row}$ that the user leaves unsatisfied, and (3) if a topic is attractive then they begin scanning items of that row with TCM behavior (i.e. with left-right cascade examination and termination probability $\kappa$ after unattractive items). By setting $\kappa_{row}=\kappa$\footnote{This is done in the formulation and the experiments presented by the authors \cite{rahdari_ranked_2022,rahdari_towards_2024}. We replicate this in the experiments below.}, the formulation simplifies to the TCM, but with less $\kappa$ terms:
\begin{equation}
P(click \mid i,j, u) = (1-\kappa)^{i+j-2}\prod_{i^{\prime}=1}^{i-1}\prod_{j^{\prime}=1}^{J}(1-\theta_{i^{\prime},j^{\prime}}) \prod_{j^{ \prime}=1}^{j-1}(1-\theta_{i,j^{\prime}}) \times \theta_{u} 
\end{equation}
The reduction in $1-\kappa$ terms is due to the TCM only modeling a user that examines items in one topic. This behavior of only examining items in one carousel row/topic is a limitation of the CCM that is not in line with eye tracked browsing behaviors seen in carousel interfaces \cite{10.1145/3742413.3789166}, where users often scanned items of multiple carousel rows. This discrepancy provided the motivation for designing and testing carousel PBMs that may perform better than the CCM.

\section{Proposal: Extending PBM to Carousels}
In this section, we explore the extension of the PBM to carousel interfaces and formally define three new carousel PBM click models, particularly the first click model to integrate eye tracking data to transform latent examination to be observable. 
In contrast to the CCM, the PBM provides the advantage of learning examination for each position separately, which allows it to effectively fit to any browsing behavior, particularly the F-pattern seen in carousels \cite{RecGaze}.

\heading{Carousel Position-Based Model (CPBM)} We naturally extend the PBM to carousel interfaces by adjusting position examination to account for rows and columns and call it the CPBM. Let $w_{i,j}$ be the probability of examining a certain position $(i,j)$. Thus, the formulation for the  CPBM can be written as:
\begin{equation}
\label{eq:Formulation Carousel PBM}
P(click \mid i,j, u) = w_{i,j} \times \theta_{u} 
\end{equation}
\heading{Carousel Row-Column PBM (RCPBM)} Rather than learn the examination probabilities of all positions independently, it may be beneficial to break down the carousel into its main elements: rows and columns. Motivated by the importance carousel rows/topics have on browsing and the F-pattern browsing  behavior in carousels \cite{10.1145/3742413.3789166,RecGaze}, we propose the RCPBM:
\begin{equation}
P(click \mid i,j, u) = w_iw_j\times \theta_{u} 
\end{equation}
where $w_i$ is the probability of examining any item in row $i$ and $w_j$ is the probability of examining any item in column $j$. 
The RCPBM has less examination variables than the CPBM, but captures within row and within column examination dependencies. 

\heading{Observed Examination PBM (OEPBM)} 
For the PBM, CPBM, and RCPBM, attraction is learned from the observable signals of clicks, while examination is inferred. It is often ideal to learn from an observable signal, as inferring latent variables can lead to poor fit (as shown in our experiments below). Eye tracking provides an observable signal from which to learn examination. The formulation for the OEPBM is the same as for the CPBM (Eq. \ref{eq:Formulation Carousel PBM}). However, the challenge lies in how to implement and fit the OEPBM with clicks and eye tracking. We show how this is done in the following section.

\section{Implementation of Click Models}
In this section, we implement click models for carousel interfaces using click feedback data. We present the first implementations of the OEPBM, the first fully observable position-based click model, as well as the first implementations of the CPBM and its row-column extension RCPBM. Additionally, we provide standardized re-implementations of the TCM and the CCM, fit directly on click data using likelihood-based optimization.\footnote{The prior implementation of the TCM and CCM relied on parameter fitting by grid search and evaluated models using total variation distance of click probabilities. Our re-implementation uses maximum likelihood estimation and log-likelihood evaluation for consistency across models.}

\heading{Problem Setup}
We consider a carousel interface composed of multiple rows and columns. Each user interaction session corresponds to a single screen containing a fixed set of items arranged in a row-column grid. The training dataset consists of tuples:
\begin{gather}
\label{eq: dataset}
        D = \bigl\{(s_t,u_t,i_t,j_t,c_t,e_t) \mid t \in \mathcal{D}\bigr\}
\end{gather} 
where $t$ is the index and $\mathcal{D}$ is the index set $\{1, 2,...,|D|\}$ of the dataset, $s$ a session, $u$ an item, $(i,j)$ its row–column position, $c \in \{0,1\}$ a click indicator, and $e \in \{0,1\}$ an examination indicator. Item attractiveness is parameterized by $\theta_u$, while examination is parametrized by $w_{i,j}$. Our objective is to estimate model parameters that maximize the log-likelihood of the observed data.

\heading{Likelihoods} We maximize two log-likelihoods: one for observed clicks and another for observed clicks and examinations. Detailed derivations of both can be found in the appendix \ref{Derivation of Likelihoods}. 
The observed clicks log-likelihood is the standard click log-likelihood:
\begin{equation}
\label{eq:LL just click}
    \mathcal{L}\mathcal{L}=\sum_{t \in \mathcal{D}} c_t \log(w_{t}\theta_t ) + (1-c_t) log(1-w_t \theta_t),
\end{equation}
using simplified notation $w_t \equiv w_{i_t,j_i}$ and $\theta_t \equiv \theta_{u_t}$.
The OEPBM requires a different log-likelihood with both observed clicks and observed examinations, which we call the observed examination log-likelihood (OELL) :
\begin{multline}
\label{eq: LL click examination}
    \mathcal{L}\mathcal{L}
    =\sum_{t \in \mathcal{D}} c_t  \log(w_{t}\theta_t )
    + (1-c_t)e_t \log(w_t (1-\theta_t))\\
    + (1-c_t)(1-e_t) \log(1-w_t)
\end{multline}
The click likelihood evaluates whether the clicks and non-clicks are predicted at the correct position and item. The OELL evaluates whether the clicks and non-clicks along with the examinations and non-examinations are predicted at the correct position and item.

\heading{Optimization Approaches}
We present three different methods for optimizing the above likelihoods: maximum likelihood estimation (MLE), expectation-maximization (EM), and gradient ascent (GA) on the likelihood. GA  is a universal approach, while MLE is used for estimation of models with observed variables, and the EM algorithm is used when there are latent variables. MLE and EM serve as classic baselines for comparison with GA. 

\subsection{TCM and CCM Fit by MLE}
Both the TCM and CCM are cascade models, which do not have examination variables. Only attraction $\theta_u$ is learned from observed clicks. Since there are no latent variables, we can simply use the MLE for parameter estimation. By clicks being distributed according to a Bernoulli (see \cref{Proof MLE}), the resulting MLE estimation for $\theta_u$ is the sample mean of click events: 
\begin{equation}
\label{MLE of clicks}
    \theta_u =\frac{\sum_{t \in \mathcal{D}_u}c_t}{|\mathcal{D}_u|}
\end{equation}
for all indices $t$ that correspond to item $u$, $\mathcal{D}_u  : \{t \in \mathcal{D}\mid u_t=u\}$.

\subsection{CPBM Fit by EM}
Unlike cascade models, the CPBM models examination as a latent parameter $w_{i,j}$ for each row-column position $(i,j)$ independently.  As there is a latent variable, the EM algorithm is used. We use the EM algorithm update rules of the standard PBM \cite{chuklin_markov_rijke_click_models_2015} and  adjust them to carousels. The update rule for iteration $m$ of the EM algorithm for the \textbf{attraction parameter} is:
\begin{equation}
    \theta_u^{(m+1)} = \frac{1}{|\mathcal{D}_u|}\sum_{t \in \mathcal{D}_u}c_t + (1-c_t)\frac{(1-w_t^{(m)})\theta_t^{(m)}}{1-w_t^{(m)}\theta_t^{(m)}}
\end{equation}
 and for the \textbf{examination parameter}:
\begin{equation}
    w_{i,j}^{(m+1)} = \frac{1}{|\mathcal{D}_{i,j}|}\sum_{t \in \mathcal{D}_{i,j}}c_t + (1-c_t)\frac{w_t^{(m)}(1-\theta_t^{(m)})}{1-w_t^{(m)}\theta_t^{(m)}}
\end{equation}
for all indices $t$ that correspond to position $(i,j)$, $\mathcal{D}_{i,j}  : \{t \in \mathcal{D}\mid i_t=i,j_t=j\}$.

\subsection{CPBM Fit by GA on Click Log-Likelihood}
Rather than use the EM, the joint likelihood can be maximized directly by gradient ascent (GA). From the standard click log-likelihood in \cref{eq:LL just click}, we can derive the update rule for gradient ascent (see \cref{derivation CPBM GA}) for iteration $m$  for the \textbf{attraction parameter}:
\begin{equation}
    \theta_u^{(m+1)} =  \theta_u^{(m)}+lr \frac{1}{|\mathcal{D}_u|} \sum_{t \in \mathcal{D}_u} \frac{c_t}{\theta_t^{(m)}} + (1-c_t) \frac{-w_t^{(m)}}{1-w_t^{(m)}\theta_t^{(m)}} 
\end{equation}
and for the \textbf{examination parameter}:
\begin{equation}
    w_{i,j}^{(m+1)} =  w_{i,j}^{(m)}+lr\frac{1}{|\mathcal{D}_{i,j}|} \sum_{t \in \mathcal{D}_{i,j}} \frac{c_t}{w_t^{(m)}} + (1-c_t) \frac{-\theta_t^{(m)}}{1-w_t^{(m)}\theta_t^{(m)}} 
\end{equation}
where $lr$ is the learning rate. For all GA methods, we clip all parameters to be in the range of $(1e-6, 1-1e-6)$.

\subsection{RCPBM Fit by GA on Click Log-Likelihood}
The RCPM is an extension of the CPBM that factors position examination into row and column examination variables breaking the independence of the positions found in the CPBM. The log-likelihood is the same as the  CPBM, but we replace the examination parameter with $row_tcol_t$:
\begin{equation}
    \mathcal{L}\mathcal{L}=\sum_{t \in \mathcal{D}} c_t \cdot log(row_{t}col_t\theta_t ) + (1-c_t) \cdot log(1-row_tcol_t \theta_t),
\end{equation}
using simplified notation $row_t \equiv w_{i_t}$ and $col_t \equiv w_{j_t}$. Similarly, we derive the gradient ascent update rule (see \cref{Derivation GA RCPBM}) for the \textbf{attraction parameter}:
\begin{equation}
    \theta_u^{(m+1)} =  \theta_u^{(m)}+ lr \frac{1}{|\mathcal{D}_u|} \sum_{t \in \mathcal{D}_u} \frac{c_t}{\theta_t^{(m)}} + (1-c_t) \frac{-row_t^{(m)} col_t^{(m)}}{1-row_t^{(m)} col_t^{(m)}\theta_t^{(m)}} 
\end{equation}
and the \textbf{row examination parameter}:
\small
\begin{equation}
    row_i^{(m+1)} =  row_i^{(m)}+ lr \frac{1}{|\mathcal{D}_i|} \sum_{t \in \mathcal{D}_i} \frac{c_t}{row_t^{(m)}} + (1-c_t) \frac{-\theta_t col_t^{(m)}}{1-row_t^{(m)} col_t^{(m)}\theta_t^{(m)}} 
\end{equation}
\normalsize
and the \textbf{column examination parameter}:
\small
\begin{equation}
    col_j^{(m+1)} =  col_j^{(m)}+lr \frac{1}{|\mathcal{D}_j|} \sum_{t \in \mathcal{D}_j} \frac{c_t}{col_t^{(m)}} + (1-c_t) \frac{-\theta_t row_t^{(m)}}{1-row_t^{(m)} col_t^{(m)}\theta_t^{(m)}} 
\end{equation}
\normalsize

\subsection{OEPBM Fit by MLE }
Rather than inferring examination, eye tracking allows learning examination from an observable signal. A position is examined if it contains one or more fixations in a session. As all variables are observable in the OEPBM, the MLE can be used. The MLE of the \textbf{attraction parameter} is the same as the TCM and CCM, the sample  mean of the click events in \cref{MLE of clicks}. Using that examination (similar to clicks) is distributed according to a Bernoulli, the MLE of the \textbf{examination parameter} is the sample mean of the examination events for a position $(i,j)$ (see \cref{MLE of clicks}):
\begin{equation}
    w_{i,j} =\frac{\sum_{t \in \mathcal{D}_{i,j}}e_t}{|\mathcal{D}_{i,j}|}
\end{equation}
%\subsection{OEPBM Fit by GA on Likelihood}
\subsection{OEPBM Fit by GA on Observed Examination Log-Likelihood}
The OEPBM uses the OELL \cref{eq: LL click examination} where clicks and examination are both observed. We derive the update rule for gradient ascent (see  \cref{Derivation GA OEPBM}) for the \textbf{attraction parameter}:
\begin{equation}
    \theta_u^{(m+1)} =  \theta_u^{(m)}+lr \frac{1}{|\mathcal{D}_u|} \sum_{t \in \mathcal{D}_u} \frac{c_t}{\theta_t^{(m)}} -  \frac{(1-c_t)e_t}{1-\theta_t^{(m)}}
\end{equation}
and the \textbf{examination parameter}:
\begin{equation}
    w_{i,j}^{(m+1)} =  w_{i,j}^{(m)}+lr \frac{1}{|\mathcal{D}_{i,j}|} \sum_{t \in \mathcal{D}_{i,j}}\frac{c_t}{w_t^{(m)}} +  \frac{(1-c_t)e_t}{w_t^{(m)}}-   \frac{(1-c_t)(1-e_t)}{1-w_t^{(m)}}
\end{equation}
\section{Experimental Methodology}
Herein, we describe the methodology of the experiments performed to be able to compare the performance of the above model implementations and answer the research questions posed. 

\heading{RecGaze Dataset} The RecGaze\footnote{Accessible at: \url{https://zenodo.org/records/15270518}} dataset~\cite{RecGaze} is the only available eye tracking and user interaction (clicks, cursor movements, and selection explanations) dataset for carousel interfaces. The data was collected in a user study, in which participants were eye tracked on desktop during 30 free-browsing screens (pick any movie on the screen). The free-browsing dataset is composed of  87 users and 30 movie selection screens totaling 2,610 screens. 47 were men and 39 were women with the following ages: (18--19): 1, (20--29): 50, (30--39): 29, (40--49): 6, and (60--69): 1. 61 participants were from Bratislava while 26 were from Amsterdam. 

The user study interface was designed to mimic the Netflix homepage with 10 carousel genre rows (Action, Animation, Comedy, Crime, Drama, Fantasy, Horror, Romance, Sci-Fi, and Thriller) each with 15 movies  ranked by popularity (number of votes) according to the IMDB non-commercial dataset. Each carousel presents 5 movies at a time, with swipe buttons to show two additional swipe sets of 5 movies. In the appendix, we include a sample visualization taken from the dataset paper \cite{RecGaze} (see \cref{fig:user_study}), while the RecGaze GitHub provides a sample gif screen recording\footnote{\url{https://github.com/santideleon/RecGaze_Dataset}}. Across all screens, the genres were randomized to account for genre preference impacting browsing/clicking behavior. 

Gaze (raw eye tracking) data was collected using a  Tobii 4C remote bar eye tracker and pre-processed into fixation data (see \cite{RecGaze} for more detail). Fixations are stable, fixed gaze points that are linked to cognitive processes of examining an item. Fixations are included in the dataset with labels corresponding to a movie poster and movie id. We use these fixations to determine if a user examines a movie. Concretely, a user examines a movie if there is at least one fixation on the movie poster during the session. %Along with gaze and fixations, the dataset includes the clicks, movie IDs, and positions of the movies. 

\heading{Creation of Click Dataset} Using the dataset, we created a click dataset as described in \cref{eq: dataset}.  It is organized by sessions, or a unique user and one of the thirty screens they interact with, each containing 10 carousels with 15 items each. We kept only sessions with a click or movie selection resulting in 2,432 sessions and clicks (i.e. users only selected/clicked one movie). In our experiments below, fixation data was used for initializations and evaluation. As a result, 45 (1.9\%) sessions with no fixations were removed. All movie ids that were never clicked were removed, as it is not possible to learn the attraction. The resulting dataset had 1,307 unique items (movies) and 2,387 clicks. We  make the dataset publicly available on Zenodo for future works in carousel click modeling.\footnote{\url{https://doi.org/10.5281/zenodo.20508189}}

\heading{Negative Sample Filtering by Impressions} For each session the dataset contains 1 click and close to 149 non-clicks. To reduce the ratio of negative samples (non-clicks) to positive samples, only the impressed (appeared on the screen/viewport) items are kept. This reduces the negative samples by almost 50\%. Clicked items are always impressed and movies are always impressed in sets of 5 (i.e. on swiping the 5 movies are replaced by a new set of 5). 

\begin{figure}
    \centering
    \includegraphics[width=1\columnwidth]{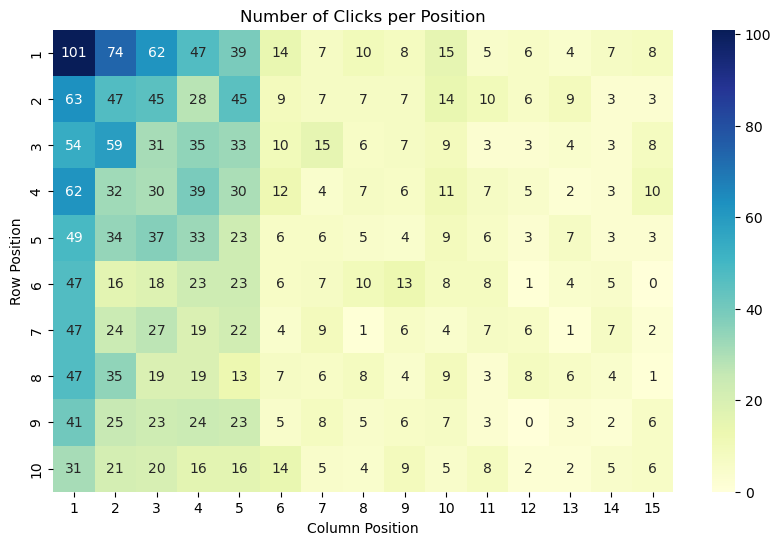} 
    \caption{Row-column position click distribution for the entire dataset. The majority of clicks are found on the first unswiped set of items (column positions 1-5) and the initial top of the page carousel rows (row positions 1-3).}
    \Description{Row-column position click distribution for the entire dataset in the form of 10 $\times$ 15 matrix. The majority of clicks are found on the first unswiped set of items (column positions 1-5) and the initial top of the page carousel rows (row positions 1-3).}
    \label{fig:row_col_click}
\end{figure}

\begin{figure}
    \centering
    \includegraphics[width=1\columnwidth]{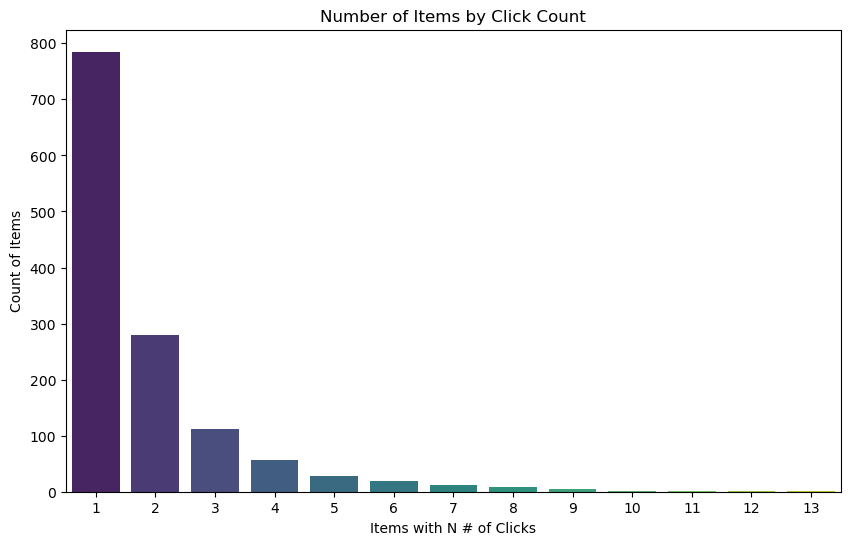} 
    \caption{Movie item click frequency for the entire dataset. The majority of items have one click, which can make it difficult to learn an attraction probability.}
    \Description{Bar chart showing the movie item click frequency for the entire dataset. The majority of items have one click, which can make it difficult to learn an attraction probability.}
    \label{fig:movie_click_freq}
\end{figure}

\begin{figure}
    \centering
    \includegraphics[width=1\columnwidth]{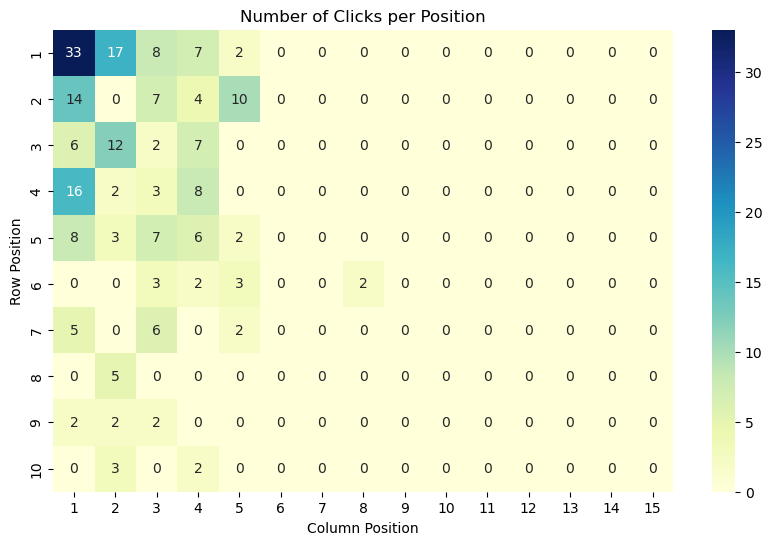} 
    \caption{Row-column position click distribution for the test set. The combination of the sparsity of multiple clicks on items and sparsity of multiple clicks on row-column positions leads to a test set that is sparse in covering the item positions of the carousel. Note that there are only 2 clicks at position (6, 8) in the test set for swiped positions.}
    \Description{Row-column position click distribution for the test set in the form of 10 $\times$ 15 matrix. The combination of the sparsity of multiple clicks on items and sparsity of multiple clicks on row-column positions leads to a test set that is sparse in covering the item positions of the carousel. Note that there are only 2 clicks at position (6, 8) in the test set for swiped positions.}
    \label{fig:row_col_click_test}
\end{figure}

\heading{Train Test Split} To create a test split, we set aside sessions of movies with 5 or more clicks and  randomly sampled half of the sessions rounding down for the test set (i.e. an item with 5 clicks has 2 test samples).  This provides a 90/10 train test split, with a test set that only contains items that have at least 3 or more clicks in the train set for learning attraction. We also maintained at least one click for each row-column position in the train set. The click distribution on row-column positions can be seen in \cref{fig:row_col_click} and Movie click frequency in \cref{fig:movie_click_freq}. The position click distribution of the test set can be seen in \cref{fig:row_col_click_test}. There are only 2 clicks in the test set for swiped positions. We comment on this sparsity in the Limitations \cref{Limitations}.

For tuning hyperparameters, the train set was split into 82/8 train and validation sets by randomly sampling half of the sessions of movies clicked at least 4 times. All final models were trained on the total train set (90\%).

\heading{Experiment Scenarios: Standard vs. Fixed Attraction} We performed two different model fitting scenarios. The first is the \textit{Standard} approach for click modeling, where models were trained with all parameters to fit the clicks. This presents two problems: (1) learning the attraction of the items and (2) learning the examination of the positions. Poor model performance can possibly be attributed to learning poor attraction values, while the examination values are correct (or vice-versa). For further comparison, we performed a \textit{Fixed Attraction} experiment, in which the attraction variable of all models was fixed to the CTR (click through rate) of the train set, as a good enough estimate, only allowing updates to examination. We hypothesize that models that perform well in the \textit{Fixed Attraction} experiment will learn better examination values more in line with user behavior.

\heading{Hyperparameters} Following the methodology for the CCM and TCM \cite{rahdari_towards_2024,rahdari_ranked_2022}, we validate termination probabilities [0.01,1] with a step size of 0.01. For gradient ascent, we validate learning rates   \{0.001, 0.01, 0.1\}. The best performing hyperparameters were decided by the best log-likelihood on the validation set.  \cref{tab:Experiment Methodology} provides a summary of the configurations of both experiments. 

\heading{Initializations and Iterations} Both GA and EM require initialization and a set amount of iterations. For item attraction we used two different initializations: uniform (0.5) and CTR. For an item, CTR was calculated on the train set taking the number of clicks divided by the number of sessions, where it was impressed. The CTR initialization is equivalent to the MLE solution for attraction, the sample mean of the click events. 

For examination we used two different initializations: Gaze and Carousel. The Gaze initialization is similar to CTR, but instead uses the number of times a position was examined. Similarly, the Gaze initialization is the equivalent to the MLE solution of examination for the OEPBM. The Carousel initialization is a rough estimate of how users browse based on prior eye tracking  \cite{10.1145/3742413.3789166}. The first row has an examination probability of 1 and all subsequent rows are cumulatively discounted by 0.95 (e.g, row 3 is $0.95^2$). Additionally, there is a single swiping discounting factor (0.7) that applies once to the second (items 5-10) and third set of items (11-15). This swiping factor is only applied once as eye tracking results \cite{10.1145/3742413.3789166} show that the first swipe is the most costly (i.e. users that swipe are likely to swipe again). For example, the position $(7,13)$ would have an examination initialization of $0.95^6(0.7)=0.51$. 

Models fit by EM and GA had all 4 possible initializations. They ran for a total of 100 iterations/epochs and were evaluated on the test set at the following iterations: 0 (initialization), 50, 100. 

\heading{Evaluation Metrics} To allow comparison, all models were evaluated with session normalized log-likelihood using both likelihoods: the click log-likelihood (Eq. \ref{eq:LL just click}) and the observed examination log-likelihood (OELL; Eq. \ref{eq: LL click examination}). The former evaluates how well the model predicts clicks/no-clicks at the correct position and item. The latter additionally evaluates whether the model correctly predicts if the position was examined or not. This leads to the OELL penalizing models with poor, unrealistic examination behaviors. In other words, the models have to reflect the actual eye tracked examination leading to examination patterns that are aligned to user behavior.

\begin{table}
    \centering
    \caption{Configurations for both experiments.}
    \label{tab:Experiment Methodology}
    \begin{tabular}{ccllc}\toprule
         Model&  Optimizer& \# Init&Eval Iter&Hyperparameters\\\midrule
         TCM&  MLE&   -&-&termination: [0.01,1]\\
         CCM&  MLE&   -&-&termination: [0.01,1]\\
 CPBM& EM&  4&\{0, 50, 100\}&-\\
 CPBM& GA&  4&\{0, 50, 100\}&lr: \{0.001, 0.01, 0.1\}\\
 RCPBM& GA&  4&\{0, 50, 100\}&lr: \{0.001, 0.01, 0.1\}\\ 
 OEPBM& MLE&  -&-&-\\
 OEPBM& GA&  4&\{0, 50, 100\}&lr: \{0.001, 0.01, 0.1\}\\ \bottomrule
    \end{tabular}
\end{table}

\section{Experimental Results}
\label{results}

\heading{Description of Results} Two experiments were performed on the impression dataset. The \textit{Standard} (learning both attraction and examination) experiment results are shown in \cref{tb: Impressed - Standard}. The \textit{Fixed Attraction} experiment results are presented in \cref{tb: Impressed - Fixed}. For each configuration of a model, algorithm, and initializations, only the best performing iteration is shown by its corresponding test log-likelihood (only the OEPBM uses OELL). Any ties are deferred to the lower iteration value. As mentioned above, the \textit{Fixed Attraction} experiment fixes the attraction to the CTR initialization (equivalent to the MLE solution). Thus, all MLE methods have the same performance across both experiments. We use the ``1\% Click'' as a baseline dummy model. For the click log-likelihood (Test LL), it always assigns 1\% to a click and 99\% to a non-click. For the observed examination log-likelihood (Test OELL), it always assigns 1\% to click and examination, 1\% to non-click and examination, and 98\% to non-click and non-examination. For the OELL, the TCM and CCM use the cascade assumption for examination values, which results in near zero values that cause the OELL to be undefined. In the following analyses, we examine general trends of models consistently outperforming across configurations/experiments, as small numerical gaps in log-likelihood are typical \cite{10.1007/978-3-319-24027-5_7}.

\begin{table}[t]
\centering
\caption{Standard Experiment Results.}
\label{tb: Impressed - Standard}
\resizebox{\columnwidth}{!}{%
\begin{tabular}{llllllll}
\toprule
    Model & Alg &   Lr & Att Init & Exam Init & Iter &   Test LL $\uparrow$ &   Test OELL $\uparrow$ \\
\midrule
1\% Click &   - &    - &        - &         - &    - &          -0.3226 &          -0.9686 \\
      TCM & MLE &    - &        - &         - &    - &          -0.2363 &                - \\
      CCM & MLE &    - &        - &         - &    - &          -0.2347 &                - \\
     CPBM &  EM &    - &      CTR &  Carousel &   50 &          -0.2304 &          -0.8515 \\
     CPBM &  EM &    - &      CTR &      Gaze &   50 &          -0.2304 &          -0.7438 \\
     CPBM &  EM &    - &  Uniform &  Carousel &   50 &          -0.2304 &          -0.9035 \\
     CPBM &  EM &    - &  Uniform &      Gaze &   50 &          -0.2304 &          -0.7720 \\
     CPBM &  GA & 0.01 &      CTR &  Carousel &   50 &          -0.2301 &          -0.8534 \\
     CPBM &  GA & 0.01 &      CTR &      Gaze &  100 &          -0.2304 &          -0.7435 \\
     CPBM &  GA & 0.01 &  Uniform &  Carousel &  100 &          -0.2321 &          -0.8647 \\
     CPBM &  GA & 0.01 &  Uniform &      Gaze &  100 &          -0.2288 &          -0.7729 \\
    RCPBM &  GA & 0.01 &      CTR &  Carousel &   50 &          -0.2300 &          -0.8170 \\
    RCPBM &  GA & 0.01 &      CTR &      Gaze &  100 &          -0.2304 &          -0.7657 \\
    RCPBM &  GA & 0.01 &  Uniform &  Carousel &  100 &          -0.2320 &          -0.9189 \\
    RCPBM &  GA &  0.1 &  Uniform &      Gaze &   50 &          -0.2304 &          -1.0146 \\
    OEPBM &  GA & 0.01 &      CTR &  Carousel &  100 &          -0.2312 &          -0.7441 \\
    OEPBM &  GA & 0.01 &      CTR &      Gaze &   50 &          -0.2312 &          -0.7440 \\
    OEPBM &  GA & 0.01 &  Uniform &  Carousel &   50 & \textbf{-0.2264} &          -0.7402 \\
    OEPBM &  GA & 0.01 &  Uniform &      Gaze &   50 &          -0.2266 & \textbf{-0.7393} \\
    OEPBM & MLE &    - &        - &         - &    - &          -0.2373 &          -0.7505 \\
\bottomrule
\end{tabular}}

\end{table}

\begin{table}
\centering
\caption{Fixed Attraction Experiment Results.}
\label{tb: Impressed - Fixed}
\resizebox{\columnwidth}{!}{%
\begin{tabular}{llllllll}
\toprule
    Model & Alg &    Lr & Att Init & Exam Init & Iter &      Test LL $\uparrow$ & Test OELL $\uparrow$ \\
\midrule
1\% Click &   - &     - &        - &         - &    - &          -0.3226 &          -0.9686 \\
      TCM & MLE &     - &        - &         - &    - &          -0.2363 &                - \\
      CCM & MLE &     - &        - &         - &    - &          -0.2347 &                - \\
     CPBM &  EM &     - &      CTR &  Carousel &    0 &          -0.2312 &          -0.8539 \\
     CPBM &  EM &     - &      CTR &      Gaze &  100 &          -0.2344 &          -0.8033 \\
     CPBM &  GA & 0.001 &      CTR &  Carousel &    0 &          -0.2312 &          -0.8539 \\
     CPBM &  GA &   0.1 &      CTR &      Gaze &  100 &          -0.2348 &          -0.7883 \\
    RCPBM &  GA &   0.1 &      CTR &  Carousel &  100 & \textbf{-0.2310} &          -0.9265 \\
    RCPBM &  GA &   0.1 &      CTR &      Gaze &  100 &          -0.2369 &          -0.7601 \\
    OEPBM &  GA &   0.1 &      CTR &  Carousel &   50 &          -0.2373 &          -0.7507 \\
    OEPBM &  GA &   0.1 &      CTR &      Gaze &    0 &          -0.2373 & \textbf{-0.7505} \\
    OEPBM & MLE &     - &        - &         - &    - &          -0.2373 & \textbf{-0.7505} \\
\bottomrule
\end{tabular}}

\end{table}

\subsection{RQ 1: Does GA Outperform EM \& MLE?}
We begin by comparing the performance of GA and MLE using the results of the OEPBM from both experiments. In the \textit{Standard} experiment, all GA initializations outperform MLE on the corresponding likelihood OELL (and on the click log-likelihood). In the \textit{Fixed Attraction} experiment, GA performs equally to MLE with the CTR and Gaze Initializations (equivalent to the MLE solutions) and performs negligibly worse (difference of 0.0002) with CTR and Carousel. These experiments support that GA can be a more effective approach than MLE for fitting click models. This is especially true when GA is initialized with the MLE solution, allowing it to further improve. However, results from the \textit{Standard} experiment (Uniform and Gaze initializations) show that exploring other initializations may lead to even better performances on both likelihoods.

Next, we compare GA and EM using the results of the CPBM. In the \textit{Standard} experiment, GA achieves the best test likelihood performance with the Uniform and Gaze initialization (same initialization as the overall best performing OEPBM). Across both experiments, GA often matches EM with certain initializations, but there are 2 cases where it performs worse, \textit{Standard}: Uniform \& Carousel; \textit{Fixed Attraction}: Gaze, with the latter only marginally under-performing (a difference of 0.0004). While more future experimental comparisons should be performed, these results point to GA being on par with EM and in some cases better.

\heading{Results Support GA can Outperform Both MLE and EM} The results above are in line with recent work \cite{hager2025claxfastflexibleneural} that suggests GA can outperform the classic methods of EM and MLE. With respect to efficiency, MLE is the fastest as it completed in one step, but GA methods can be implemented very efficiently, especially compared to EM \cite{hager2025claxfastflexibleneural}. For fitting carousel click models, GA with MLE solution initializations along with other initializations (which may provide even better results) appear to be a strong and reliable approach. 
 
\subsection{RQ 2: Do PBM Variants Better Model Carousel Clicks Than  Cascade Models?}
For these analyses, we only examine the Test LL or standard click log-likelihood, \cref{eq:LL just click}. Under the click log-likelihood, a model is not penalized for an incorrect examination as long as the product of examination and attraction provides the correct click/non-click. 

With regards to the TCM and CCM, our results are in line with prior works \cite{rahdari_ranked_2022,rahdari_towards_2024} that show the CCM consistently outperforms the TCM. Comparing cascade models to PBMs, for the \textit{Standard} experiment, all carousel PBM variants (across all initializations) outperform the CCM, except for one case: \textit{Standard} OEPBM fit by MLE with a difference of 0.0026. Note that both MLE and GA approaches of the OEPBM optimize the OELL. It is reasonable that the under-performing MLE (as compared to GA) can lead to a worse performance on the non-objective click likelihood. For the \textit{Fixed Attraction} experiment, the CCM outperforms the OEPBM and only one configuration of the CPBM by a marginal 0.0001 difference. Outside of a few cases of cascade models outperforming the OEPBM, PBM variants perform better on the standard likelihood across both experiments when compared to the TCM and CCM.

When comparing PBM variants among themselves, in the \textit{Standard} experiment the majority of PBMs perform similarly around -0.2304, which as mentioned before outperforms the CCM and TCM. However, a few outperforming outliers can be seen in the \textit{Standard} experiment with the OEPBM GA initializations of Uniform and Carousels (1st: -0.2264) and initializations of Uniform and Gaze (2nd: -0.2266). In the \textit{Fixed Attraction} experiment, the RCPBM with initialization CTR and Carousel outperforms all other models (1st: -0.2310), but the CPBM GA with initializations CTR and Carousel is a close second (2nd: -0.2312). These results may possibly be attributed to the  Carousel initialization for examination being more suited to the \textit{Fixed Attraction} experiment, as all models performed equal or better with this configuration than with other initializations. 

\heading{PBM Variants are Better Than Cascade Baselines and OEPBM is the Best} Considering a standard click modeling approach, results point to the OEPBM outperforming other models with two consistent results as the best and second best. When additionally considering the observed examination likelihood, there is even more of an argument for picking the OEPBM (which we further discuss in the next section). However, the OEPBM requires gathering observable examination data. In the cases of not having examination data, both the CPBM or RCPBM can be used with GA optimizer, a CTR initialization for attraction, and an examination initialization that provides some prior knowledge of how users examine items (e.g, positional bias from click feedback). 

\subsection{RQ 3: Do Carousel PBMs Reflect Realistic Examination Behavior?}
For these final analyses, we jointly examine both the click log-likelihood and the observed examination log-likelihood (OELL). Note that the OELL is a metric that is better aligned with user behavior, as the model must correctly estimate examinations and clicks. This is a significantly harder problem, leading to poorer values for the OELL compared to the Test LL. Click likelihood can be seen as a binary classification problem, while observed likelihood is a 4-class problem. This 4-class problem is not solvable by the TCM and CCM. The problem lies in that the TCM and CCM are models that are capable of modeling clicks through mathematical optimization, but with examination probabilities that are infeasible when faced with a more realistic metric.  

Examining both experiments, a pattern emerges: models that have an examination initialization of Gaze consistently have better OELL. Most importantly this is true for models that are not trained with the OELL objective (i.e. CPBM and RCPBM). In the \textit{Standard} experiment, the CPBM and RCPBM with Gaze initializations have similar click likelihood performances as the Carousel initialization, but the Carousel initializations show some of the worst performances on the OELL. Based on these results, initializing examination with examination data can lead to a model that does not sacrifice click likelihood and has much more realistic examination values.

Focusing on the OEPBM, it consistently performs well on the OELL and always outperforms the CPBM and RCPBM, but this is not surprising as it is the objective likelihood. In line with the results of the CPBM and RCPBM, the Gaze initialization across both experiments provides the best OELL.

\begin{figure*}
    \centering
    \includegraphics[width=0.85\textwidth]{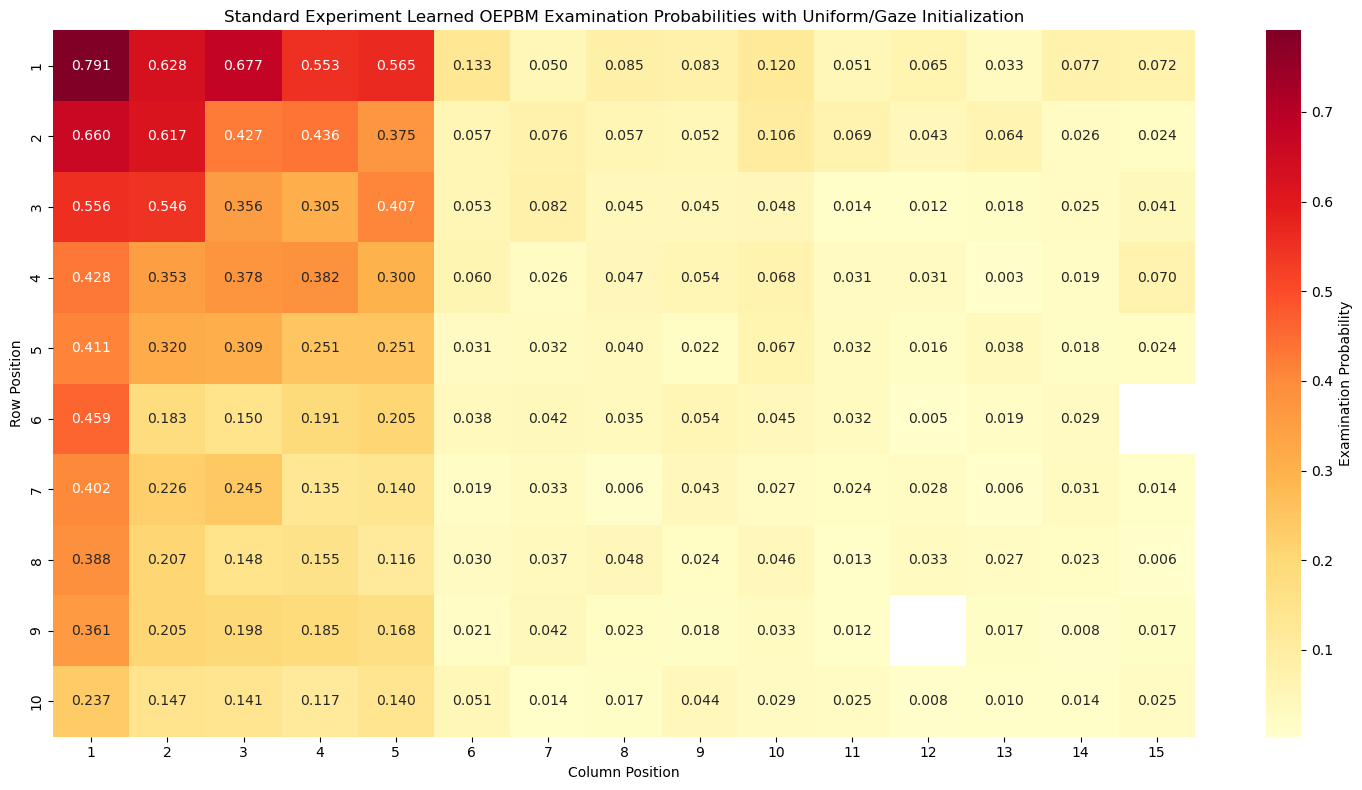} 
    \caption{Examination probabilities of the first, unswiped page (columns 1-5) follow the standard F-pattern browsing behavior of carousel users. For pages 2 (columns 6-10) and pages 3 (columns 11-15) that are swiped to, the OEPBM learns the general trend of the mirrored F-pattern (right to left) despite the sparse click data for these positions, resulting in lower probabilities. OEPBM learns examination probabilities that are in line with realistic user behavior.}
    \Description{A matrix of 10 rows $\times$ 15 columns. Examination probabilities of the first, unswiped page (columns 1-5) follow the standard F-pattern browsing behavior of carousel users. For pages 2 (columns 6-10) and pages 3 (columns 11-15) that are swiped to, the OEPBM learns the general trend of the mirrored F-pattern (right to left) despite the sparse click data for these positions, resulting in lower probabilities. OEPBM learns examination probabilities that are in line with realistic user behavior.}
    \label{fig:exam_prob}
\end{figure*}

To further validate the OEPBM's ability to model realistic examination behavior, we analyzed the learned examination probabilities. \cref{fig:exam_prob} shows the position examination values learned by the best performing OELL model configuration in the standard experiment: OEPBM optimized with gradient ascent with initializations Uniform and Gaze. Eye tracking results \cite{RecGaze,10.1145/3742413.3789166} have shown that users follow an F-pattern examination behavior on the initial, unswiped items and after swiping follow a mirrored F-pattern from right to left. On the initial, unswiped set of items (columns 1-5) the OEPBM learns the F-pattern with higher values tending to the top-left. For the second (columns 6-10) and third (columns 11-15) swiped set of items, a general trend following the mirrored F-pattern can be seen with the higher values tending to the top-right. The lower values for second and third swiped sets are attributed to the sparse clicks found on these positions. Despite the sparsity of data, the OEPBM is able to capture the mirrored F-pattern behavior after swiping.

\heading{OEPBM is the Most Realistic and Click Likelihood can be Misaligned with Realistic Behavior} Following from research question 2, the OEPBM not only fits best to the click likelihood, but also the OELL. This leads to the OEPBM to better reflect user examination behavior, as it learns from the examination data. A way to impart this to latent examination models is through a gaze-informed initialization (i.e. transfer learning), which does not harm click likelihood performance. However, to maximize click likelihood and OELL, it appears that more data than clicks is needed. This points to an important general insight for click models: click models that are fit to click likelihoods do not guarantee a good fit to user browsing behavior, especially for complex interfaces. While model design can help improve click likelihoods and be more realistic, especially when compared to  cascade models, the only remedy to Goodhart's Law, in this case, is more information on realistic user behavior provided by eye tracking.  

\section{Limitations}
\label{Limitations}
The main limitation of this work is the dataset. The RecGaze Dataset \cite{RecGaze} is the only available dataset that combines click feedback with eye tracking data (in general and) for carousel interfaces. It contains 1,307 unique items and 2387 clicks, making it much smaller than common industrial click datasets that can have hundreds of thousands of clicks or more. Additionally, there is the problem of item click sparsity (see \cref{fig:movie_click_freq}), which is also common to other click datasets. This can make it difficult to correctly determine attraction values for items with little clicks, which can also impact examination values. We address this by guaranteeing the train set has at least 3 clicks for each item in the test. Despite this, the dataset still presents a difficult scenario for the click models to learn appropriate attraction and examination values. Even with this difficulty, all models outperform the dummy model baseline. 

In terms of position, click position sparsity is also present in the dataset (see \cref{fig:row_col_click,fig:row_col_click_test}). This is to be expected, as users are much more likely to click items on the first page. The evaluation of the click models fit on this dataset may not reflect behaviors of users that are searching deeper into swiped lists due to the lack of data. However, as was seen with the Gaze initialization applied to the PBM variants, the eye tracking data is very rich and can provide valuable information for examination even when there are no clicks on those positions. This is due to examination being a much more common and less sparse action than clicks. While limitations are present, it is reasonable to have such limitations with a non-industrial dataset. Furthermore, interface-aware click models are underdeveloped. While our PBM formulations are generalizable across varying carousel interfaces (i.e. different number of rows/columns and swiping behaviors), the empirical fit of the models may not transfer as well, especially across domains. We hope to encourage further work in developing interface-aware click models and that further experiments be done with larger industrial datasets to validate our results in many settings. 

Another limitation is the general lack of eye tracking data in click modeling or recommendation settings. While large-scale eye tracking is not yet standard for industry, emerging technologies (e.g., mobile front-facing cameras already capable of eye-tracking, AR/VR) make this increasingly plausible. Other observable signals (e.g., cursor dwell time, scrolling behavior) can also be used to enrich modeling, while aggregate click data (especially with debiasing methods  \cite{10.1145/3018661.3018699}) can serve to approximate examination. Rather than large-scale eye tracking, smaller eye tracking user studies can be conducted to allow the use of the OEPBM more as a validation tool for verifying production user models that lack gaze data. 

\section{Conclusion}

In this work, we studied click modeling for carousel interfaces, which, despite their prevalence in e-commerce, are underexplored. We explored both model design and optimization strategies and proposed three novel carousel PBMs. Through our experiments, we showed that gradient-based optimization methods outperform classic approaches in fitting carousel click models. Among the evaluated models, OEPBM achieved the strongest performance in click prediction and learned examination parameters that most closely align to user behavior. Additionally, our results reveal a fundamental limitation of click modeling in complex interfaces: optimizing for click likelihood does not guarantee realistic modeling of user examination. Clicks alone do not provide sufficient information, requiring additional behavioral feedback, such as eye tracking. 

These findings point to the importance of user feedback outside of clicks and interface-aware approaches in recommender systems. Future work involves adapting other single-list click models to carousel (or other) interfaces, developing new click models that may better capture the complexity of user behavior (e.g. implementing neural click models), or integrating other sources of user feedback (e.g., gaze, cursor, dwell time, etc.). Along with improving click model alignment to user behavior,  further work is needed exploring how these models can improve downstream evaluation metrics or the quality of the recommendation for the user. 

\section{Ethical Use of Data}
The publicly available RecGaze Dataset \cite{RecGaze} was used following the terms of use and licensing. All participants provided informed consent for participation and creation of the dataset.

\begin{acks}
This work was supported by Eyes4ICU, a project funded by the European Union under the Horizon Europe Marie Sk\l{}odowska-Curie Actions, GA No. \href{https://doi.org/10.3030/101072410}{101072410} and by LorAI - Low Resource Artificial Intelligence, a project funded by the European Union, GA No. \href{https://doi.org/10.3030/101136646}{101136646}. 

\end{acks}

\bibliographystyle{ACM-Reference-Format}
\balance
\bibliography{references.bib}

\appendix
\begin{figure*}
    \centering
    \includegraphics[width=0.8\textwidth]{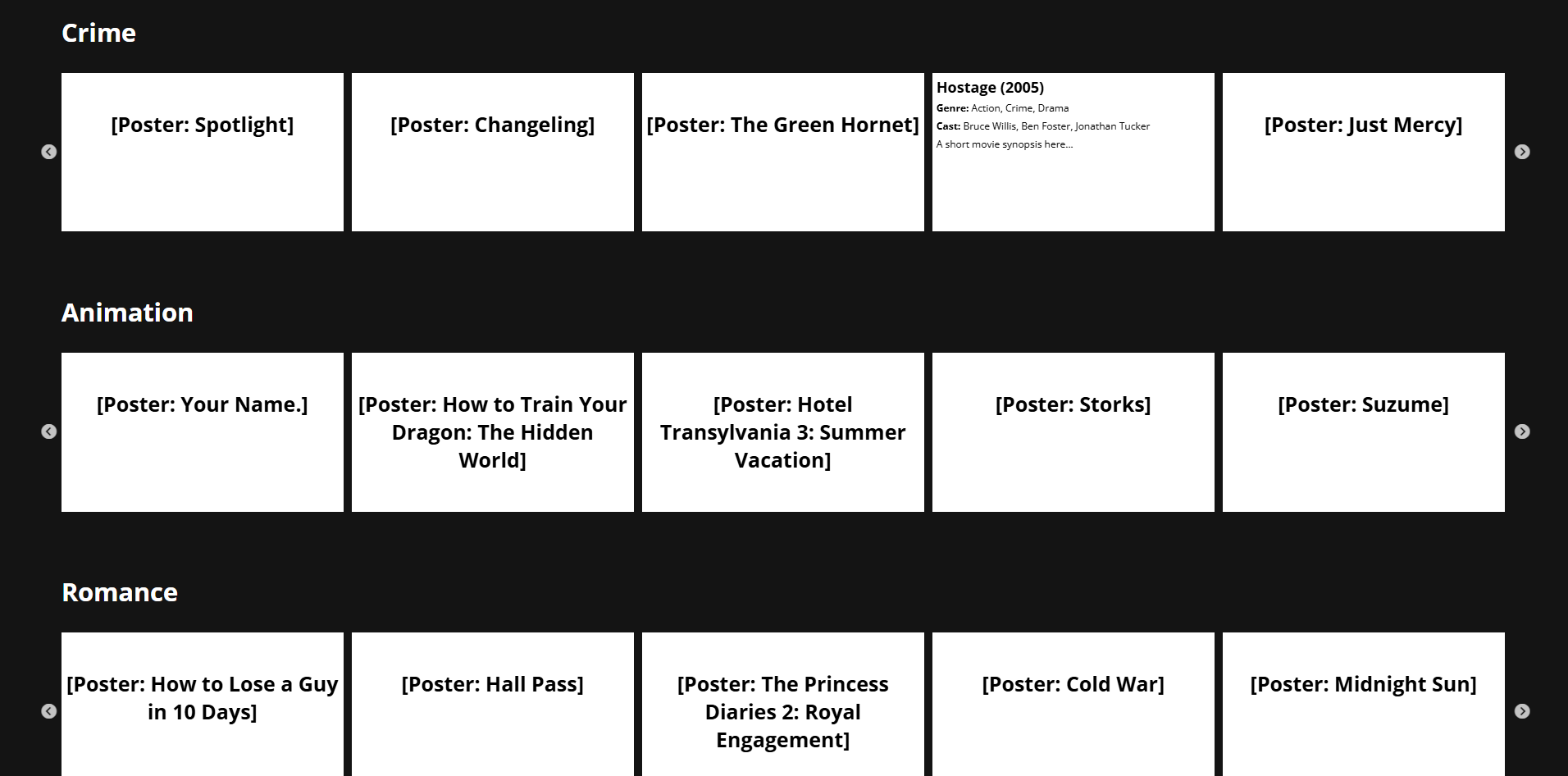}
    \caption{A sample screen of a carousel interface from the user study as presented in~\cite{RecGaze}. It captures an initial presentation with the first 3 carousels shown and mouse-over details of a movie in the top row. Movie posters and mouse-over details were shown  in the actual study and are removed here only due to the copyright.}
    \Description{A sample screen of a carousel interface from the user study as presented in~\cite{RecGaze}. It captures an initial presentation with the first 3 carousels shown and mouse-over details of a movie in the top row (at position (1, 4)). Movie posters and mouse-over details were shown in the actual study and are removed here only due to the copyright.}
    \label{fig:user_study}
\end{figure*}

\section{Derivation of Likelihoods}
\label{Derivation of Likelihoods}
Let C be a random variable for clicks that is Bernoulli distributed,  $C\sim Bernoulli(\Theta)$, where $\theta$ is the attraction parameter. Let E be a random variable for examination that is Bernoulli distributed, $E\sim Bernoulli(W)$, where $w$ is the examination parameter. Attraction is defined for each item $u$  of the form $\theta_u=P(\Theta_u=1)$. Examination is defined for each row-column position $(i,j)$ of the form $w_{i,j}=P(E_{i,j}=1)$. All of the mentioned click models follow the same basic assumption: \textit{for a click to happen, the position/item must be examined and the item found to be attractive.}

By the above assumption, we can separate the probability of a click into two cases: (1) clicked and (2) not clicked:
\begin{align}
\label{eq:prob just click app}
    P(C=1 \mid i,j,u) = P(E_{i,j}=1)P(\Theta_u=1 ) =w_{i,j}\theta_u\\
    P(C=0 \mid i,j,u) = 1- P(C=1 \mid i,j,u)  = 1-w_{i,j}\theta_u
\end{align}
The above is sufficient to define a log-likelihood for observed clicks from $D$ (see \cref{eq: dataset}) evaluated over the indices $t$:
\begin{equation}
\label{eq:LL just click app}
    \mathcal{L}\mathcal{L}=\sum_{t \in \mathcal{D}} c_t \log(w_{t}\theta_t ) + (1-c_t) \log(1-w_t \theta_t),
\end{equation}
where $c_t$ is the click event at index $t$ that is either clicked ($c_t=1)$ or not clicked $(c_t=0)$ and using simplified notation $w_t \equiv w_{i_t,j_i}$ and $\theta_t \equiv \theta_{u_t}$.

If we consider the joint probability of clicks $C$ and examination $E$, then we have four different cases that are possible rather than the two above: (1) clicked and examined, (2) clicked and not examined, (3) not clicked and examined, and (4) not clicked and not examined. Note that for case (2) by the above assumptions there cannot be a click and no examination.\footnote{In the data there are no cases of a click without an observation and even if there were, it would be a good idea to remove them (either it is an error in the examination data or it is a random click)} Also by the above assumption, we know the item is attractive in (1), not attractive in (3), and in (4) we don't know if it is attractive or not. Using the four cases, the joint probability of a click and examination can be written as the following:
\small
\begin{gather}
\label{eq: prob click examination app}
    P(C=1, E=1\mid i,j,u) = P(E_{i,j}=1)P(\Theta_u=1 ) =w_{i,j}\theta_u\\
    P(C=1, E=0 \mid i,j,u) = 0\\
    P(C=0, E= 1\mid i,j,u) = P(E_{i,j}=1)P(\Theta_u=0) =  w_{i,j}(1-\theta_u)\\
    P(C=0, E= 0\mid i,j,u) =P(E_{i,j}=0) = (1-w_{i,j})
\end{gather}
\normalsize
Now we define the log-likelihood for observed clicks and observed examination, without case (2) clicked and not examined:
\begin{gather}
\label{eq: LL click examination app}
    \mathcal{L}\mathcal{L}
    =\sum_{t \in \mathcal{D}} c_te_t \log(w_{t}\theta_t )
    + (1-c_t)e_t \log(w_t (1-\theta_t))\\
    + (1-c_t)(1-e_t) \log(1-w_t)
\end{gather}
where $e_t$ is the examination event for index $t$ that is either examined ($e_t=1)$ or not examined $(e_t=0)$, effectively acting as a indicator along with $c_t$. We can further simplify by using the assumption that all clicked items are examined ($c_te_t=c_t$).

\section{Derivation of MLE for TCM and CCM}
\label{Proof MLE}
First, we derive the likelihood of $\theta_u$. Using that attraction is learned from the observable clicks distributed according to a Bernoulli distribution $C  \sim Bernoulli(\Theta)$, we can derive the likelihood for all indices $t$ that correspond to item $u$, $\mathcal{D}_u  : \{t \in \mathcal{D}\mid u_t=u\}$, of the following form:
\begin{equation}
    \mathcal{L}(\theta_u) = \prod_{t \in \mathcal{D}_u} \theta_t^{c_t} (1-\theta_t)^{1-c_t}
\end{equation}
and the log-likelihood:
\begin{equation}
    \mathcal{L}\mathcal{L}(\theta_u) =  \sum_{t \in \mathcal{D}_u} c_t  \log(\theta_t)+ (1-c_t)  \log(1-\theta_t)
\end{equation}
To get the MLE estimation of the parameter, we now take its derivative and set it zero:
\begin{gather*}
        0 =  \sum_{t \in \mathcal{D}_u} c_t\frac{ 1}{\theta_t}+ (1-c_t) \frac{-1}{1-\theta_t}\\
       0= \sum_{t \in \mathcal{D}_u}c_t(1-\theta_t)-\theta_t(1-c_t)\\
        \sum_{t \in \mathcal{D}_u}c_t=\sum_{t \in \mathcal{D}_u}\theta_t-\theta_t c_t+\theta_t c_t\\
         \sum_{t \in \mathcal{D}_u}c_t=\sum_{t \in \mathcal{D}_u}\theta_t\\
         \sum_{t \in \mathcal{D}_u}c_t=|\mathcal{D}_u|\theta_u \\
        \theta_u =\frac{\sum_{t \in \mathcal{D}_u}c_t}{|\mathcal{D}_u|}
\end{gather*}
The resulting MLE estimation of $\theta_u$ is the sample mean of the click events.

\section{Derivation of GA for CPBM}
\label{derivation CPBM GA}
The log-likelihood of the CPBM with a observed clicks and latent examinations is \cref{eq:LL just click app} above:
\begin{equation}
    \mathcal{L}\mathcal{L}=\sum_{t \in \mathcal{D}} c_t  \log(w_{t}\theta_t ) + (1-c_t)  \log(1-w_t \theta_t),
\end{equation}
For gradient ascent on the likelihood we will need both the partial derivatives of the attraction parameter and the examination parameter. The partial derivative of the attraction parameter is:
\begin{equation}
    \frac{\partial \mathcal{L}\mathcal{L}}{\partial \theta_t} = \frac{c_t}{\theta_t} + (1-c_t) \frac{-w_t}{1-w_t\theta_t}
\end{equation}
and for the examination parameter:
\begin{equation}
    \frac{\partial \mathcal{L}\mathcal{L}}{\partial w_t} = \frac{c_t}{w_t} + (1-c_t) \frac{-\theta_t}{1-w_t\theta_t}
\end{equation}
Thus, the update rule for gradient ascent on iteration $m$  for the attraction parameter is:
\begin{equation}
    \theta_u^{(m+1)} =  \theta_u^{(m)}+lr \frac{1}{|\mathcal{D}_u|} \sum_{t \in \mathcal{D}_u} \frac{c_t}{\theta_t^{(m)}} + (1-c_t) \frac{-w_t^{(m)}}{1-w_t^{(m)}\theta_t^{(m)}} 
\end{equation}
The update rule for the examination parameter is:
\begin{equation}
    w_{i,j}^{(m+1)} =  w_{i,j}^{(m)}+lr  \frac{1}{|\mathcal{D}_{i,j}|} \sum_{t \in \mathcal{D}_{i,j}} \frac{c_t}{w_t^{(m)}} + (1-c_t) \frac{-\theta_t^{(m)}}{1-w_t^{(m)}\theta_t^{(m)}} 
\end{equation}
for all indices $t$ that correspond to position $(i,j)$, $\mathcal{D}_{i,j}  : \{t \in \mathcal{D}\mid i_t=i,j_t=j\}$ and where $lr$ is the learning rate.

\section{Derivation of GA for RCPBM }
\label{Derivation GA RCPBM}
The log-likelihood is the same as the Carousel PBM, but we replace the examination parameter with $row_tcol_t$:
\begin{equation}
    \mathcal{L}\mathcal{L}=\sum_{t \in \mathcal{D}} c_t  \log(row_{t}col_t\theta_t ) + (1-c_t)  \log(1-row_tcol_t \theta_t),
\end{equation}
using simplified notation $row_t \equiv w_{i_t}$ and $col_t \equiv w_{j_t}$. Similarly we take the partial derivatives with respect to each variable. For the attraction parameter:
\begin{equation}
    \frac{\partial \mathcal{L}\mathcal{L}}{\partial \theta_t} = \frac{c_t}{\theta_t} + (1-c_t) \frac{-row_tcol_t}{1-row_tcol_t\theta_t}
\end{equation}
and the row examination parameter:
\begin{equation}
    \frac{\partial \mathcal{L}\mathcal{L}}{\partial row_t} = \frac{c_t}{row_t} + (1-c_t) \frac{-\theta_t col_t}{1-row_t col_t\theta_t}
\end{equation}
and the column examination parameter:
\begin{equation}
    \frac{\partial \mathcal{L}\mathcal{L}}{\partial col_t} = \frac{c_t}{col_t} + (1-c_t) \frac{-\theta_t row_t}{1-row_t col_t\theta_t}
\end{equation}
Thus, the gradient ascent update rule for the attraction parameter is:
\begin{equation}
    \theta_u^{(m+1)} =  \theta_u^{(m)}+lr \frac{1}{|\mathcal{D}_u|} \sum_{t \in \mathcal{D}_u} \frac{c_t}{\theta_t^{(m)}} + (1-c_t) \frac{-row_t^{(m)} col_t^{(m)}}{1-row_t^{(m)} col_t^{(m)}\theta_t^{(m)}} 
\end{equation}
The update rule for the row examination parameter is evaluated for all indices $t$ that correspond to row $i$, $\mathcal{D}_{i}  : \{t \in \mathcal{D}\mid i_t=i\}$:
\small
\begin{equation}
    row_i^{(m+1)} =  row_i^{(m)}+lr  \frac{1}{|\mathcal{D}_i|} \sum_{t \in \mathcal{D}_i} \frac{c_t}{row_t^{(m)}} + (1-c_t) \frac{-\theta_t col_t^{(m)}}{1-row_t^{(m)} col_t^{(m)}\theta_t^{(m)}} 
\end{equation}
\normalsize
The update rule for the column examination parameter is evaluated for all indices $t$ that correspond to column $j$, $\mathcal{D}_{j}  : \{t \in \mathcal{D}\mid j_t=j\}$:
\small
\begin{equation}
    col_j^{(m+1)} =  col_j^{(m)}+lr  \frac{1}{|\mathcal{D}_j|} \sum_{t \in \mathcal{D}_j} \frac{c_t}{col_t^{(m)}} + (1-c_t) \frac{-\theta_t row_t^{(m)}}{1-row_t^{(m)} col_t^{(m)}\theta_t^{(m)}} 
\end{equation}
\normalsize

\section{Derivation of GA for OEPBM}
\label{Derivation GA OEPBM}
The OEPBM uses the second log-likelihood \cref{eq: LL click examination} above where clicks and examination are both observed:  
\begin{gather}
    \mathcal{L}\mathcal{L}
    =\sum_{t \in \mathcal{D}} c_t  \log(w_{t}\theta_t )
    + (1-c_t)e_t \log(w_t (1-\theta_t))\\
    + (1-c_t)(1-e_t) \log(1-w_t)
\end{gather}
We take the partial derivatives with respect to each variable. For the attraction parameter:
\begin{equation}
    \frac{\partial \mathcal{L}\mathcal{L}}{\partial \theta_t} = \frac{c_t}{\theta_t} -  \frac{(1-c_t)e_t}{1-\theta_t}
\end{equation}
For the examination parameter:
\begin{equation}
    \frac{\partial \mathcal{L}\mathcal{L}}{\partial w_t} = \frac{c_t}{w_t} +  \frac{(1-c_t)e_t}{w_t}-   \frac{(1-c_t)(1-e_t)}{1-w_t}
\end{equation}
Now we derive the update rule for gradient ascent for the attraction parameter and iteration $m$:
\begin{equation}
    \theta_u^{(m+1)} =  \theta_u^{(m)}+lr \frac{1}{|\mathcal{D}_u|} \sum_{t \in \mathcal{D}_u} \frac{c_t}{\theta_t^{(m)}} -  \frac{(1-c_t)e_t}{1-\theta_t^{(m)}} 
\end{equation}
and the gradient ascent update rule for the examination parameter:
\begin{equation}
    w_{i,j}^{(m+1)} =  w_{i,j}^{(m)}+lr  \frac{1}{|\mathcal{D}_{i,j}|} \sum_{t \in \mathcal{D}_{i,j}}\frac{c_t}{w_t^{(m)}} +  \frac{(1-c_t)e_t}{w_t^{(m)}}-   \frac{(1-c_t)(1-e_t)}{1-w_t^{(m)}}
\end{equation}

\end{document}